\documentclass[12pt,preprint]{aastex}

\usepackage{amsmath}
\usepackage{graphicx}
\usepackage{comment}
\usepackage{color} 
\newcommand{\fig}[1]{Fig.~\ref{#1}}

\newcommand{\sect}[1]{Sect.~\ref{#1}}
\newcommand{\equ}[1]{Equation~(\ref{#1})}

\begin{document}
\title{Confined and eruptive catastrophes of solar magnetic flux ropes caused by mass loading and unloading}
\author{Quanhao Zhang\altaffilmark{1,2,3}, Rui Liu\altaffilmark{1,2,4}, Yuming Wang\altaffilmark{1,2,3}, Xiaolei Li\altaffilmark{1,3}, Shaoyu Lyu\altaffilmark{1,3}}
\altaffiltext{1}{CAS Key Laboratory of Geospace Environment, Department of Geophysics and Planetary Sciences, University of Science and Technology of China, Hefei 230026, China}
\altaffiltext{2}{CAS Center for Excellence in Comparative Planetology, University of Science and Technology of China, Hefei 230026, China}
\altaffiltext{3}{Mengcheng National Geophysical Observatory, School of Earth and Space Sciences, University of Science and Technology of China, Hefei 230026, China}
\altaffiltext{4}{Collaborative Innovation Center of Astronautical Science and Technology, Hefei, Anhui 230026, China}
\email{zhangqh@ustc.edu.cn}

\begin{abstract}
It is widely accepted that coronal magnetic flux ropes are the core structures of large-scale solar eruptive activities, which inflict dramatic impacts on the solar-terrestrial system. Previous studies have demonstrated that varying magnetic properties of a coronal flux rope system could result in a catastrophe of the rope, which may trigger solar eruptive activities. Since the total mass of a flux rope also plays an important role in stabilizing the rope, we use 2.5-dimensional magnetohydrodynamic (MHD) numerical simulations in this letter to investigate how a flux rope evolves as its total mass varies. It is found that an unloading process that decreases the total mass of the rope could result in an upward (eruptive) catastrophe in the flux rope system, during which the rope jumps upward and the magnetic energy is released. This indicates that mass unloading processes could initiate the eruption of the flux rope. Moreover, when the system is not too diffusive, there is also a downward (confined) catastrophe that could be caused by mass loading processes, via which the total mass accumulates. The magnetic energy, however, is increased during the downward catastrophe, indicating that mass loading processes could cause confined activities that may contribute to the storage of energy before the onset of coronal eruptions.
\end{abstract}


\section{Introduction}
\label{sec:introduction}
Large-scale solar eruptions include filament/prominence eruptions, coronal mass ejections (CMEs), and flares. They are intimately related to each other, and are usually believed to be different manifestations of coronal magnetic flux rope eruptions \citep{Zhang2001,Vrvsnak2005a,Jiang2016a,Yan2020,Liu2020a}. Since large-scale solar eruptions are the major trigger of extreme space weather  \citep{svestka2001a,Cheng2014,Lugaz2017,Gopalswamy2018a}, it is significant to investigate the eruptive mechanism of coronal flux ropes. To shed light on the physical processes of solar eruptive activities, many efforts have been made in modeling the eruptions of coronal flux ropes in previous theoretical studies, in which distinctive physical mechanisms are invoked, e.g., ideal magnetohydrodynamic(MHD) instabilities  \citep{Torok2003a,Aulanier2010a,Guo2010,Savcheva2012b,Keppens2019}, magnetic reconnections  \citep{Antiochos1999a,Chen2000a,Moore2001a,Sterling2004,Archontis2008b,Inoue2015}, and flux rope catastrophes. 
\par
The flux rope catastrophe theory investigates the onset of flux rope eruptions from the perspective of loss of equilibrium. Before the onset of the eruption, the flux rope should be in equilibrium (static or quasi-static), i.e., with no net force on the rope; after the onset of the eruption, however, the flux rope evolves in a dynamic process, during which the state of motion must be very different from that before the onset. The varying state of motion implies that a loss of equilibrium occurs at the onset of eruption, so that the net force resulting from the loss of equilibrium initiates the eruption of the rope. This is the physical scenario of flux rope catastrophes, which was first proposed by \cite{vanTend1978a}: they analytically derived the equilibrium manifold as a function of the filament current, and found that the catastrophe occurs if the current exceeds a critical value, which is called the catastrophic point. Here the equilibrium manifold is the set of all the equilibrium states of the flux rope system; based on the equilibrium manifold, one is able to tell when and how the catastrophe occurs. Similar results were obtained by many other analytical studies \citep[][]{Forbes1991a,Lin2000a,Demoulin2010a,Longcope2014a,Kliem2014}, suggesting that the catastrophes of coronal flux ropes are promising candidates for coronal eruptions.
\par
Numerical simulations are also widely used to investigate flux rope catastrophes \citep[e.g.,][]{Forbes1990a,Chen2007a,Zhang2017a}. In those studies, the authors calculated equilibrium solutions of the rope system with different values of a characteristic physical parameter, and all the equilibrium solutions constitute the equilibrium manifold versus this parameter. For example, by simulating the equilibrium manifold as a function of the magnetic field strength within the flux rope, \cite{Hu2001a} concluded that a catastrophe occurs if the strength exceeds a critical value. Similar conclusions are reached in many other studies \citep[e.g.,][]{Su2011a,Zhang2017a}. In most of the previous studies, the flux rope jumps upward when the catastrophe occurs, so that this kind of catastrophe is termed ``upward catastrophe''. Recently, \cite{Zhang2016a} found its downward counterpart caused by reducing magnetic fluxes of coronal flux ropes; during this catastrophe, the flux rope suspended in the corona falls down to the photosphere. This new kind of catastrophe is thus termed ``downward catastrophe'', and \cite{Zhang2016a} also demonstrated that the magnetic energy is released during not only the upward but also the downward catastrophe. Observational studies suggested that, apart from coronal eruptions, there are also various confined activities, with no portion of the flux rope escaping the corona \cite[e.g.][]{Gilbert2000,Jiang2016a}. Comparing with the observations, \cite{Zhang2016a} inferred that the downward catastrophe could be responsible for, e.g., confined flares.
\par
Previous numerical studies mainly focus on flux rope catastrophes associated with the magnetic parameters of the rope; whether the variation of the mass could cause flux rope catastrophes, however, has hardly been touched upon previously. In fact, as concluded by \cite{Low1996a}, the total mass of a flux rope should play an important role in stabilizing the rope, which is further confirmed by some recent studies \citep{Hillier2013,Jenkins2019,Tsap2019,Fan2020}. Moreover, the total mass of a flux rope is usually not conserved but highly dynamic \citep{Zhang2017b,Gibson2018}: it either accumulates via, e.g., coronal condensation \citep{Liu2012d,Xia2016,Jenkins2021}, or decreases as a result of, e.g., mass unloading \citep{Low1996a} and mass drainage processes \citep{Bi2014,Zhang2017b,Jenkins2019}. Therefore, to understand the influence of the varying mass on the catastrophic behaviors of coronal flux ropes is not only important to improve flux rope catastrophe theory, but also shed light on the physical scenario of different kinds of flux rope activities. To achieve this, we carry out numerical simulations to seek for the flux rope catastrophes associated with the rope mass. For simplicity, we use ``mass loading'' and ``mass unloading'' to refer to the processes that increase and decrease the mass hereafter, respectively. The rest of this letter is arranged as follows. The simulating procedures are introduced in \sect{sec:model}. The simulations are presented in \sect{sec:result}. Our conclusion and discussion are given in \sect{sec:dc}.

\section{Simulating procedures}
\label{sec:model}
\begin{figure*}
\includegraphics[width=1.0\hsize]{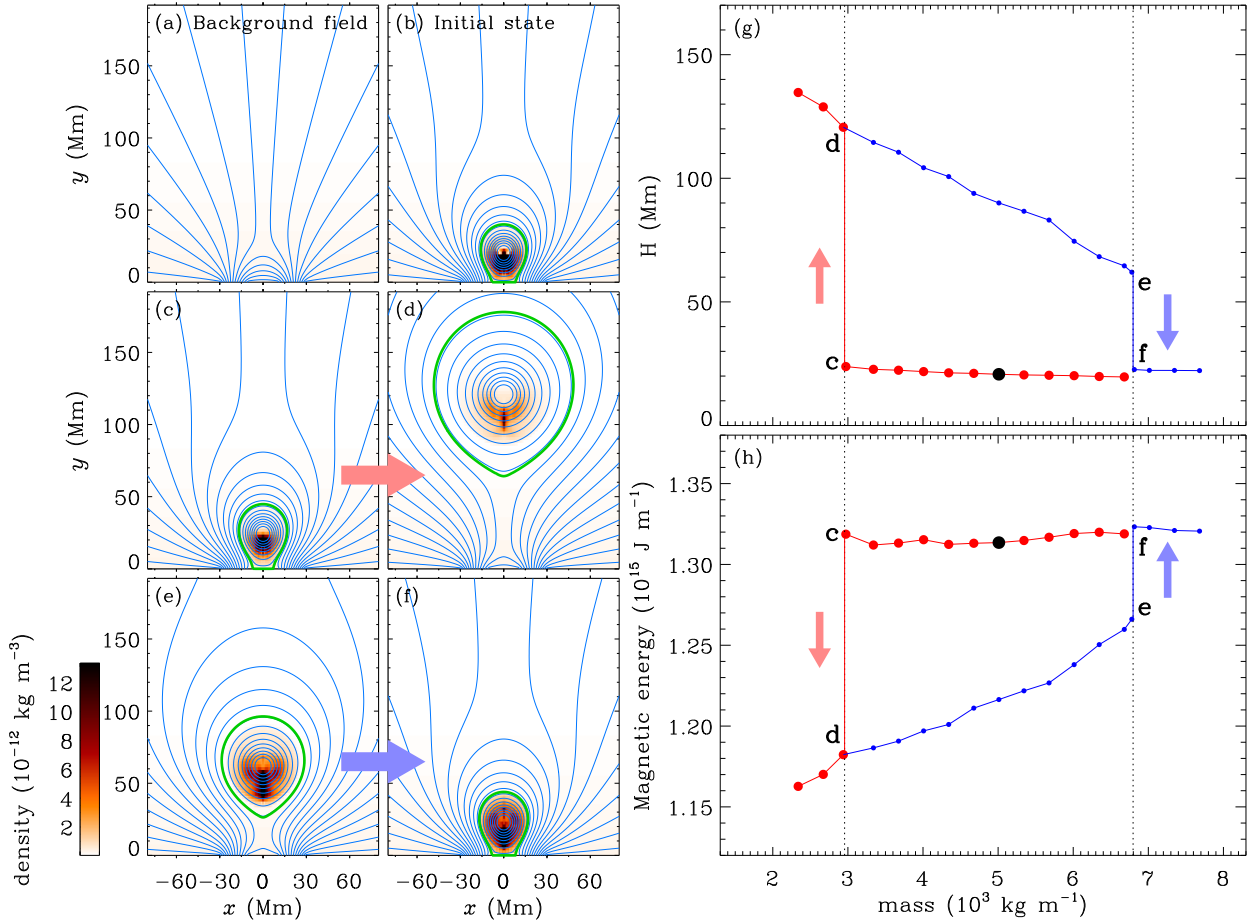}
\caption{Simulations of the catastrophes caused by the varing mass of the rope. Panel (a) is the magnetic configuration of the background field, and panel (b) that of the initial state; panels (c) and (d) are the equilibrium states right before and after the upward catastrophe, and panels (e) and (f) are those for the downward catastrophe. In panels (a)-(f), the blue lines represent the magnetic field lines; the green curve mark the boundary of the flux rope; the red color depicts the distribution of the mass density. The dots in panel (g) plot the variation of the height of the rope axis for the simulated equilibrium states as a function of the rope mass, and the corresponding magnetic energies of the equilibrium states are plotted by the dots in panel (h). The upward (eruptive) and the downward (confined) catastrophes are marked by the red and blue arrows, respectively.}\label{fig:result}
\end{figure*}

In our 2.5-dimensional simulations, all the quantities satisfy $\partial/\partial z=0$. The magnetic field can be written as:
\begin{align}
\textbf{B}=\triangledown\psi\times\hat{\textbf{\emph{z}}}+B_z\hat{\textbf{\emph{z}}},\label{equ:mf}
\end{align}
Here $\psi$ is the magnetic flux function; the subscript $z$ denotes the component in $z-$dirention. Basic equations and numerical procedures to construct the initial state are introduced in Appendix \ref{apdx}. The magnetic configuration of the background field is plotted in \fig{fig:result}(a): it is a partially open bipolar field, with a negative and a positive surface magnetic charges located at the lower base within $-30~\mathrm{Mm}<x<-10~\mathrm{Mm}$ and $10~\mathrm{Mm}<x<30~\mathrm{Mm}$, respectively. The initial state, illustrated in \fig{fig:result}(b), is an equilibrium state consisting of a flux rope embedded in the background field. Anomalous resistivity is used:
\begin{align}
\eta=
\begin{cases}
0,& ~j\leq j_c\\
\eta_m\mu_0v_0L_0(\frac{j}{j_c}-1)^2.& ~j> j_c \label{equ:res}
\end{cases}
\end{align}
Here $\eta_m=10^{-4}$, $L_0=10^7$ m, $v_0=128.57$ km s$^{-1}$, and $j_c=2.37\times10^{-4}$ A m$^{-2}$; $\mu_0$ is the vacuum magnetic permeability. The physical properties of a flux rope can be characterized by its axial magnetic flux, $\Phi_z$, its poloidal magnetic flux per unit length along the $z$-direction, $\Phi_p$, and its total mass per unit length along the $z$-direction, $M$. For the flux rope in the initial state, the values of the these properties are: $\Phi_{z0}=4.10\times10^{19}~\mathrm{Mx}$, $\Phi_{p0}=1.49\times10^{12}~\mathrm{Mx}~\mathrm{m}^{-1}$, and $M_0=5.01\times10^3$ kg m$^{-1}$. Assuming the length of the flux rope to be 100 Mm, its total mass is of the order $M_0=5.0\times10^{11}$ kg, which falls within the observed mass range of typical prominences/filaments \citep{Parenti2014a}. 
\par
To investigate the evolution of the rope versus its total mass, we simulate the equilibrium manifold as a function of the mass. Starting from the initial state, we first adjust the mass density at each grid within the rope by
\begin{align}
\rho=\alpha\rho_0,~\alpha=\frac{M_1}{M_0},
\end{align}
where $\rho_0$ is the density in the initial state. With this procedure, the total mass of the rope is adjusted from $M_0$ to a certain target value $M_1$. Second, we let the rope system relax to an equilibrium state, during which the total mass of the rope is maintained to be conserved at $M_1$ by similar procedures to those in \cite{Hu2003a}. Moreover, relaxation method \citep[e.g.,][]{Zhang2016a} is used in the region outside the flux rope: reset the temperature and density to their initial values, so that the pressure gradient force is always balanced by the gravitational force; when the rope system eventually evolves to the equilibrium state, force-free condition is satisfied in the region outside the flux rope. The region inside the flux rope is still non-force-free. With the simulating procedures introduced above, we obtain the equilibrium state with the rope mass to be $M_1$. Similar procedures are repeated for different target values of the mass (e.g,. $M_2$, $M_3$, ...), and all the calculated equilibrium states constitute the equilibrium manifold. During the whole simulation, $\Phi_z$ and $\Phi_p$ are always maintained to be conserved at $\Phi_{z0}$ and $\Phi_{p0}$ by similar procedures to those in \cite{Hu2003a}, i.e., the characteristic magnetic parameters of all the simulated equilibrium states are the same, so that we could focus on the influence of the mass on the rope system.
\par
In our simulation, the quantities at the lower boundary are fixed, indicating that it corresponds to the photosphere; increment equivalent extrapolation is used at the other boundaries \citep[e.g.,][]{Zhang2020}:
\begin{align*}
U^{n+1}_{b}=U^{n+1}_{b-1}+U^{n}_{b}-U^{n}_{b-1}.
\end{align*}
Here $U$ represents the quantities (e.g. $\rho$, $\textbf{v}$, $\psi$); the quantities at the current and the next time steps are indicated by the superscript n and n+1, respectively; $U_b$ represents the quantities at the boundary, and $U_{b-1}$ the quantities at the location next to the boundary. The boundary quantities at the next time step, $U^{n+1}_{b}$, are then evaluated.

\section{Simulation results}
\label{sec:result}
The simulated equilibrium manifold of the flux rope system is plotted in \fig{fig:result}(g). The evolution of the equilibrium states versus the mass is described by the variation of the height of the rope axis, $H$. The black dot in \fig{fig:result}(g) corresponds to the initial state \fig{fig:result}(b), starting from which we change the mass of the rope and calculate the corresponding equilibrium states, as plotted by the red dots. There is a critical value of the mass, $M^u=2.96\times10^3$ kg m$^{-1}$: the flux rope keeps sticking to the photosphere if its total mass $M$ is larger than $M^u$, as shown in, e.g., \fig{fig:result}(c); if $M$ decreases to reach $M^u$, however, the flux rope jumps upward, and evolves to an equilibrium state with the flux rope suspended in the corona, as shown in \fig{fig:result}(d). Consequently, the equilibrium manifold is discontinuous at $M^u$ (marked by the vertical red line in \fig{fig:result}(g)), i.e., an upward catastrophe occurs at $M^u$ if $M$ decreases to reach $M^u$, as marked by the red arrow in \fig{fig:result}(g); $M^u$ is the upward catastrophic point. \fig{fig:result}(c) and \ref{fig:result}(d) illustrate the equilibrium states right before and after the upward catastrophe occurs, respectively; they are also marked by ``c'' and ``d'' in \fig{fig:result}(g). We note that the resistivity in our simulation is not very large ($\eta_m=10^{-4}$), so that the rope does not escape from the corona after the upward catastrophe occurs, but keeps suspended in the corona (\fig{fig:result}(d)). If the system is strongly diffusive (e.g., $\eta_m=0.1$), the rope could keep rising after the upward catastrophe occurs, resulting in a full eruption. In addition, it could be recognized in \fig{fig:result}(g) that $H$ slowly increases with the decreasing $M$ before the upward catastrophe occurs, which may correspond to the slow-rising phase before the onset of the eruption. Our simulation result of the upward catastrophe caused by the decreasing mass demonstrates that mass unloading processes could initiate flux rope eruptions. This result is consistent with the analytical analysis in \cite{Jenkins2019}. Moreover, by using the analytical model in \cite{Jenkins2019}, we could estimate the critical height at which loss of equilibrium should occur in our model. The input parameters include: the mass $m=M^u=2.96\times10^3$ kg m$^{-1}$, half of the distance between the two photospheric polarities $D=10$ Mm, the magnetic field strength at the photosphere $B_{Phot}=B(x=0,y=0)=5.97$ G. The critical height is then calculated as 17.32 Mm. In our simulations, $H$ of the state right before the upward catastrophe occurs (\fig{fig:result}(c)) is 23.84 Mm. The gap between the critical heights should result from the difference in the models: the flux rope in our simulation has a finite radius, whereas a thin-rope model is used in \cite{Jenkins2019}.
\par
Restarting from the equilibrium state with the rope suspended in the corona (\fig{fig:result}(d)), we increase the total mass of the rope and simulate the corresponding equilibrium states, which are plotted by the blue dots in \fig{fig:result}(g). Obviously, the flux rope does not immediately evolve back to the state marked as ``c'' when we start to decrease the mass of the rope, but keeps suspended in the corona, as shown in, e.g., \fig{fig:result}(e). This indicates that the upward catastrophe is an irreversible process. As demonstrated by the blue dots in \fig{fig:result}(g), there is another critical mass $M^d=6.81\times10^3$ kg m$^{-1}$, at which the flux rope falls down until reaching an equilibrium state with the rope once again sticking to the photosphere (\fig{fig:result}(f)); the eventual magnetic configuration is similar to the initial state. Therefore, a downward catastrophe occurs if the mass of the rope increases to reach $M^d$. The equilibrium states right before and after the downward catastrophe, which are illustrated in \fig{fig:result}(e) and \fig{fig:result}(f), are also marked by ``e'' and ``f'' in \fig{fig:result}(g), and the downward catastrophe is marked by the blue arrow. We have also tried to decrease the mass from the equilibrium state after the downward catastrophe occurs (\fig{fig:result}(f)), and verified that the downward catastrophe is also an irreversible process, similar to the upward catastrophe; if we decrease the mass to the upward catastrophic point, $M^u$, the upward catastrophe occurs again. Our results suggest that mass loading processes could cause confined activities in flux rope systems before the onset of the eruption. From \fig{fig:result}(g), we may conclude that the equilibrium states are bifurcated into two branches: the states with the rope sticking to the photosphere belong to the lower branch, and those with the rope suspended in the corona belong to the upper branch. These two branches are separated from each other, and are only connected by the upward and the downward catastrophes. Obviously, the upward and the downward catastrophe should occur in flux rope systems with different topologies: the equilibrium state right before the upward catastrophe occurs (\fig{fig:result}(c)) is in the lower branch, belonging to the bald patch separatrix surface (BPSS) topology \citep{Titov1993a}, whereas that for the downward catastrophe (\fig{fig:result}(e)) is in the upper branch, which should correspond to the hyperbolic flux tube (HFT) topology \citep{Titov2003}. As shown in \fig{fig:result}(g), the rope will keep rising (descending) if its total mass decreases (increases) with time until reaching the catastrophic points in both the upper and the lower branches. Since faster variation of the mass corresponds to faster variation of the rope height, higher mass loss (gain) rate should result in larger rising (descending) velocity of the rope. It is noteworthy that the downward catastrophe could only exist when the magnetic system is not too diffusive; otherwise, as explicated above, the flux rope will escape the corona after the upward catastrophe occurs, so that the upper branch as well as the downward catastrophe vanishes. 
\par
To roughly estimate the variation of the magnetic energy versus the total mass of the rope, we calculate the magnetic energy within the domain for each simulated equilibrium state, which are plotted in \fig{fig:result}(h). Although the axial and poloidal magnetic fluxes of the flux ropes in all the simulated equilibrium states are the same (\sect{sec:model}), the corresponding magnetic energies are quite different. The equilibrium states belonging to the lower branch tend to have larger magnetic energies than those belonging to the upper branch; the two branches are also well separated in the parameter space spanned by the magnetic energy and the mass. This should be caused by the different topologies of the states in different branches. As a result, magnetic energy is released during the upward catastrophe, via which the flux rope system evolves from the high-energy state (lower branch) to the low-energy state (upper branch); in contrast, magnetic energy is stored during the downward catastrophe. The amount of the magnetic energy released during the upward catastrophe is of the order $E_{m}^u=13.64\times10^{10}$ J~m$^{-1}$, and that stored during the downward catastrophe is of the order $E_{m}^d=5.72\times10^{10}$ J~m$^{-1}$. Moreover, to further investigate the energy budget during the catastrophes, we also calculate the variations of the gravitational energy during the catastrophes: it increases by $E_{g}^u=6.89\times10^{10}$ J~m$^{-1}$ during the upward catastrophe, and decreases by $E_{g}^u=6.23\times10^{10}$ J~m$^{-1}$ during the downward catastrophe. Comparing with $E_{m}^u$ and $E_{m}^d$, we may infer that: (1) during the upward catastrophe, part (about 50$\%$) of the released magnetic energy is converted to the gravitational energy; (2) the stored magnetic energy during the downward catastrophe might originate from the gravitational energy in the flux rope system. Moreover, comparing \fig{fig:result}(g)) and \fig{fig:result}(h), we may infer that the increase of the magnetic energy as a result of the increasing mass in the upper branch should counter the according decrease of the rope height. This result might presumably be useful for the analysis of, e.g., quiescent prominences.

\section{Discussion and conclusion}
\label{sec:dc}
In this letter, we investigate how the varying mass of coronal flux ropes affects their catastrophic evolutions. Our simulation results discover both the upward (eruptive) and the downward (confined) catastrophes associated with the varying mass of the rope. For the flux rope system with the rope sticking to the photosphere, an upward catastrophe will occur if the mass of the rope decreases to reach a critical value. During the upward catastrophe, the flux rope jumps upward, and the magnetic energy is released. Therefore, the upward catastrophe should initiate the eruption of the flux rope. This result provides a promising theoretical scenario for triggering the eruption of coronal flux ropes in the BPSS configuration caused by mass unloading processes. Moreover, if the total mass of a flux rope suspended in the corona  increases to reach another critical value, a downward catastrophe will occur, during which the flux rope falls down to the photosphere, and the magnetic energy is, however, stored in the rope system. This indicates that mass loading processes could cause confined activities of coronal flux ropes in the HFT configuration, which may contribute to the storage of energy before the onset of solar eruptions. These two catastrophes appear as discontinuous jumps in the equilibrium manifold (as marked by the arrows in \fig{fig:result}(g)); they connect the separated upper and lower branches of the equilibrium states of the rope system, and both of the catastrophes are irreversible processes. We note once more that the resistivity in our simulation is not very large; if the magnetic system is too diffusive, the upper branch as well as the downward catastrophe vanishes, indicating that the downward catastrophe could only exist in, e.g., quiescent regions.
\par
As simulated in \cite{Zhang2016a}, increasing and decreasing magnetic fluxes are also able to cause upward and downward catastrophes, respectively; the magnetic energy, however, is released during both the upward and the downward catastrophes caused by varying magnetic fluxes. Comparing with our simulation results, we may conclude that increasing magnetic fluxes of a coronal flux rope should play a similar role as the decreasing mass of the rope: they are in favor of coronal eruptions. Nevertheless, although both decreasing magnetic fluxes and increasing mass could cause confined activities, the corresponding energetic evolutions are quite different: the downward catastrophe caused by decreasing magnetic fluxes provides a possible scenario for confined flares, whereas that caused by the increasing mass corresponds to the storage process of the magnetic energy. This implies that downward catastrophes, regulated by different physical parameters, may contribute to various confined activities observed in the solar corona. Efforts are still needed to seek for observational counterparts of these phenomena.
\par
In this letter, our analysis is mainly based on the parameter space of the mass, rather than the temporal evolution of the rope. Since there are various kinds of non-linear processes that could change the mass of the rope, the temporal evolution of the rope associated with these processes should reveal the evolutionary scenario of coronal flux ropes in the actual solar atmosphere. We will extend our model and investigate this topic in our future studies.

\begin{acknowledgements}
The authors appreciate the anonymous referee for the comments and suggestions that helped to improve this work. This research is supported by the National Natural Science Foundation of China (NSFC 41804161, 42174213, 41774178, 41761134088, 41774150, 41842037 and 41574165), the Strategic Priority Program of CAS (XDB41000000 and XDA15017300), and the fundamental research funds for the central universities. Quanhao Zhang acknowledges for the support from National Space Science Data Center, National Science \verb"&" Technology Infrastructure of China (www.nssdc.ac.cn). 
\end{acknowledgements}

\appendix
\section{Basic equations and initial preparations}
\label{apdx}
With the form in \equ{equ:mf}, the MHD equations could then be cast in the non-dimensional form as:
\begin{align}
&\frac{\partial\rho}{\partial t}+\triangledown\cdot(\rho\textbf{\emph{v}})=0,\label{equ:cal-st}\\
\nonumber &\frac{\partial\textbf{\emph{v}}}{\partial t}+\frac{2}{\rho\beta_0}(\vartriangle\psi\triangledown\psi+B_z\triangledown B_z+\triangledown\psi\times\triangledown B_z)+\textbf{\emph{v}}\cdot\triangledown\textbf{\emph{v}}\\ 
&~~~+\triangledown T +\frac{T}{\rho}\triangledown\rho+g\hat{\textbf{\emph{y}}}=0,\\
&\frac{\partial\psi}{\partial t}+\textbf{\emph{v}}\cdot\triangledown\psi-\eta\vartriangle\psi=0,\\
&\frac{\partial B_z}{\partial t}+\triangledown\cdot(B_z\textbf{\emph{v}})+(\triangledown\psi\times\triangledown v_z)\cdot\hat{\textbf{\emph{z}}}-\eta\vartriangle B_z=0,\\
\nonumber &\frac{\partial T}{\partial t}-\frac{\eta(\gamma-1)}{\rho R}\left[(\vartriangle\psi)^2+|\triangledown\times(B_z\hat{\textbf{\emph{z}}})|^2 \right]\\
&~~~+\textbf{\emph{v}}\cdot\triangledown T +(\gamma-1)T\triangledown\cdot\textbf{\emph{v}}=0,\label{equ:cal-en}
\end{align}
where
\begin{align}
\vartriangle\psi=\frac{\partial^2\psi}{\partial x^2}+\frac{\partial^2\psi}{\partial y^2},~~\vartriangle B_z=\frac{\partial^2 B_z}{\partial x^2}+\frac{\partial^2 B_z}{\partial y^2}.
\end{align}
Here $\rho$ and $T$ are the density and the temperature; $\textit{\textbf{v}}= v_x\hat{\textbf{\emph{x}}}+v_y\hat{\textbf{\emph{y}}}+v_z\hat{\textbf{\emph{z}}}$ is the velocity; $g$ is the normalized gravity, which is a constant; $\eta$ is the resistivity; $\gamma=5/3$ is the polytropic index; $\beta_0=0.1$ is the characteristic ratio of the gas pressure to the magnetic pressure. The characteristic values of the quantities are: $\rho_0=3.34\times10^{-13}\mathrm{~kg~m^{-3}}$, $T_0=10^6\mathrm{~K}$, $\psi_0=3.73\times10^3\mathrm{~Wb~m^{-1}}$, $v_0=128.57$ km s$^{-1}$, $t_0=77.8$ s, $B_0=3.37\times10^{-4}$ T, $g_0=1.65\times10^3$ m s$^{-2}$. The numerical domain is $0<x<300$ Mm, $0<y<300$ Mm; it is discretized into 300$\times$150 uniform meshes. Symmetric boundary condition is used at the left side of the domain ($x=0$). The radiation and the heat conduction in the energy equation are neglected.
\par
To investigate the catastrophic behaviors of coronal flux rope systems versus the mass of the rope, we must first construct a typical flux rope system. Here we first use the complex variable method \citep[e.g.,][]{Zhang2020} to construct the partially open bipolar background field, which can be cast in the complex variable form
\begin{align}
f(\omega)\equiv B_x-iB_y=\frac{(\omega+iy_N)^{1/2}(\omega-iy_N)^{1/2}}{F(a,b,y_N)}\mathrm{ln}\left( \frac{\omega^2-a^2}{\omega^2-b^2}\right),
\end{align}
where $\omega=x+iy$, and
\begin{align}
\nonumber &F(a,b,y_N)=\frac{1}{b-a}\int_a^b(x^2+y_N^2)^{1/2}dx=\frac{1}{2(b-a)}\times\\ &\left[b(b^2+y_N^2)^{1/2}-a(a^2+y_N^2)^{1/2}+y_N^2\mathrm{ln}\left(\frac{b+(b^2+y_N^2)^{1/2}}{a+(a^2+y_N^2)^{1/2}} \right)\right],
\end{align}
with $a=10$ Mm, $b=30$ Mm, and ($y=y_N=26.3$ Mm, $x=0$) is the position of the neutral point in the background field; a neutral current sheet is located at $(x=0,~y\geq y_N)$. The magnetic flux function $\psi$ could then be obtained by:
\begin{align}
\psi(x,y)=\mathrm{Im}\left\lbrace\int f(\omega)d\omega \right\rbrace.\label{equ:integral}
\end{align}
The flux function at the lower base ($y=0$), $\psi_i$, is
\begin{equation}
 \psi_i=\psi(x,0) = \left\{
              \begin{array}{ll}
              {\psi_c}, &{|x|<a}\\
              {\psi_c F(|x|,b,y_N)/F(a,b,y_N)}, &{a\leqslant|x|\leqslant b}\\
              {0}, &{|x|>b}
              \end{array}  
         \right.\label{equ:fluxb}
\end{equation}
where $\psi_c=2\pi\psi_0$, and that at the neutral point $y=y_N$ is:
\begin{align}
\psi_N=\frac{\pi(b^2-a^2)}{2F(a,b,y_N)}.\label{equ:fluxc}
\end{align}
With the obtained $\psi$, and letting $B_z=0$ in the background field, the distribution of the magnetic field within the domain could be calculated. The initial corona is isothermal and static:
\begin{align}
T_c\equiv T(t=0,x,y)=1\times10^6 ~\mathrm{K},\ \  \rho_c\equiv\rho(t=0,x,y)=\rho_0\mathrm{e}^{-gy}.\label{equ:rhot}
\end{align}
\par
Starting from the initial condition and the background configuration, equations (\ref{equ:cal-st}) to (\ref{equ:cal-en}) are simulated by the multi-step implicit scheme \citep{Hu1989a}. With similar simulating procedures to those in \cite{Zhang2020}, we let a flux rope emerge from the lower base at a constant speed. The emergence begins at $t=0$ and ends at $t=\tau_E=174$ s, so that at time $t$ ($0\leqslant t\leqslant \tau_e$), the emerged part of the rope is within $-x_E\leqslant x\leqslant x_E$ at the lower base, where $x_E=(a^2-h_E^2)^{1/2}$, $h_E=a(2t/\tau_E-1)$.
Thus the emergence is achieved by adjusting the quantities at the lower base ($y=0$) within $-x_E\leqslant x\leqslant x_E$ 
\begin{align}
&\psi(t,x,y=0)=\psi_i(x,y=0)+\psi_E(t,x),\\
&\psi_E(t,x)=\frac{C_E}{2}\mathrm{ln}\left(\frac{2a^2}{a^2+x^2+h_E^2}\right)\label{equ:psi},\\ 
&B_z(t,x,y=0)=C_Ea(a^2+x^2+h_E^2)^{-1}\label{equ:bz},\\
&v_y(t,x,y=0)=2a/\tau_E,~v_x(t,x,y=0)=v_z(t,x,y=0)=0,\\
&T(t,x,y=0)=2\times10^5\mathrm{~K},~\rho(t,x,y=0)=1.67\times10^{-12}\mathrm{~kg~m^{-3}},
\end{align}
where $C_E$=3.0. With these procedures, we obtain a flux rope system consisting of a magnetic flux rope embedded in the background field, as shown in \fig{fig:result}(b). We note that the emergence of the rope is merely a numerical approach to construct a coronal flux rope system, so that the detailed evolution during the emergence should not be related to the physics discussed in this paper.


\begin{thebibliography}{}
\expandafter\ifx\csname natexlab\endcsname\relax\def\natexlab#1{#1}\fi

\bibitem[{{Antiochos} {et~al.}(1999){Antiochos}, {DeVore}, \&
  {Klimchuk}}]{Antiochos1999a}
{Antiochos}, S.~K., {DeVore}, C.~R., \& {Klimchuk}, J.~A. 1999, \apj, 510, 485

\bibitem[{{Archontis} \& {Hood}(2008)}]{Archontis2008b}
{Archontis}, V., \& {Hood}, A.~W. 2008, \apjl, 674, L113

\bibitem[{{Aulanier} {et~al.}(2010){Aulanier}, {T{\"o}r{\"o}k}, {D{\'e}moulin},
  \& {DeLuca}}]{Aulanier2010a}
{Aulanier}, G., {T{\"o}r{\"o}k}, T., {D{\'e}moulin}, P., \& {DeLuca}, E.~E.
  2010, \apj, 708, 314

\bibitem[{Bi {et~al.}(2014)Bi, Jiang, Yang, Hong, Li, Yang, \& Yang}]{Bi2014}
Bi, Y., Jiang, Y., Yang, J., {et~al.} 2014, \apj, 790, 100

\bibitem[{{Chen} \& {Shibata}(2000)}]{Chen2000a}
{Chen}, P.~F., \& {Shibata}, K. 2000, \apj, 545, 524

\bibitem[{{Chen} {et~al.}(2007){Chen}, {Hu}, \& {Sun}}]{Chen2007a}
{Chen}, Y., {Hu}, Y.~Q., \& {Sun}, S.~J. 2007, \apj, 665, 1421

\bibitem[{{Cheng} {et~al.}(2014){Cheng}, {Ding}, {Zhang}, {Sun}, {Guo}, {Wang},
  {Kliem}, \& {Deng}}]{Cheng2014}
{Cheng}, X., {Ding}, M.~D., {Zhang}, J., {et~al.} 2014, \apj, 789, 93

\bibitem[{{D{\'e}moulin} \& {Aulanier}(2010)}]{Demoulin2010a}
{D{\'e}moulin}, P., \& {Aulanier}, G. 2010, \apj, 718, 1388

\bibitem[{{Fan}(2020)}]{Fan2020}
{Fan}, Y. 2020, \apj, 898, 34

\bibitem[{{Forbes}(1990)}]{Forbes1990a}
{Forbes}, T.~G. 1990, \jgr, 95, 11919

\bibitem[{{Forbes} \& {Isenberg}(1991)}]{Forbes1991a}
{Forbes}, T.~G., \& {Isenberg}, P.~A. 1991, \apj, 373, 294

\bibitem[{{Gibson}(2018)}]{Gibson2018}
{Gibson}, S.~E. 2018, Living Reviews in Solar Physics, 15, 7

\bibitem[{{Gilbert} {et~al.}(2000){Gilbert}, {Holzer}, {Burkepile}, \&
  {Hundhausen}}]{Gilbert2000}
{Gilbert}, H.~R., {Holzer}, T.~E., {Burkepile}, J.~T., \& {Hundhausen}, A.~J.
  2000, \apj, 537, 503

\bibitem[{{Gopalswamy} {et~al.}(2018){Gopalswamy}, {Akiyama}, {Yashiro}, \&
  {Xie}}]{Gopalswamy2018a}
{Gopalswamy}, N., {Akiyama}, S., {Yashiro}, S., \& {Xie}, H. 2018, Journal of
  Atmospheric and Solar-Terrestrial Physics, 180, 35

\bibitem[{{Guo} {et~al.}(2010){Guo}, {Ding}, {Schmieder}, {Li},
  {T{\"o}r{\"o}k}, \& {Wiegelmann}}]{Guo2010}
{Guo}, Y., {Ding}, M.~D., {Schmieder}, B., {et~al.} 2010, \apjl, 725, L38

\bibitem[{Hillier \& van Ballegooijen(2013)}]{Hillier2013}
Hillier, A., \& van Ballegooijen, A. 2013, \apj, 766, 126

\bibitem[{{Hu}(1989)}]{Hu1989a}
{Hu}, Y.~Q. 1989, Journal of Computational Physics, 84, 441

\bibitem[{Hu(2001)}]{Hu2001a}
Hu, Y.~Q. 2001, Solar Physics, 200, 115

\bibitem[{{Hu} {et~al.}(2003){Hu}, {Li}, \& {Xing}}]{Hu2003a}
{Hu}, Y.~Q., {Li}, G.~Q., \& {Xing}, X.~Y. 2003, Journal of Geophysical
  Research (Space Physics), 108, 1072

\bibitem[{{Inoue} {et~al.}(2015){Inoue}, {Hayashi}, {Magara}, {Choe}, \&
  {Park}}]{Inoue2015}
{Inoue}, S., {Hayashi}, K., {Magara}, T., {Choe}, G.~S., \& {Park}, Y.~D. 2015,
  \apj, 803, 73

\bibitem[{Jenkins {et~al.}(2019)Jenkins, Hopwood, D{\'e}moulin, Valori,
  Aulanier, Long, \& van Driel-Gesztelyi}]{Jenkins2019}
Jenkins, J.~M., Hopwood, M., D{\'e}moulin, P., {et~al.} 2019, \apj, 873, 49

\bibitem[{Jenkins \& Keppens(2021)}]{Jenkins2021}
Jenkins, J.~M., \& Keppens, R. 2021, \aap, 646, A134

\bibitem[{Jiang {et~al.}(2016)Jiang, Wu, Yurchyshyn, Wang, Feng, \&
  Hu}]{Jiang2016a}
Jiang, C., Wu, S.~T., Yurchyshyn, V., {et~al.} 2016, \apj, 828, 62

\bibitem[{Keppens {et~al.}(2019)Keppens, Guo, Makwana, Mei, Ripperda, Xia, \&
  Zhao}]{Keppens2019}
Keppens, R., Guo, Y., Makwana, K., {et~al.} 2019, Reviews of Modern Plasma
  Physics, 3, 14

\bibitem[{{Kliem} {et~al.}(2014){Kliem}, {Lin}, {Forbes}, {Priest}, \&
  {T{\"o}r{\"o}k}}]{Kliem2014}
{Kliem}, B., {Lin}, J., {Forbes}, T.~G., {Priest}, E.~R., \& {T{\"o}r{\"o}k},
  T. 2014, \apj, 789, 46

\bibitem[{{Lin} \& {Forbes}(2000)}]{Lin2000a}
{Lin}, J., \& {Forbes}, T.~G. 2000, \jgr, 105, 2375

\bibitem[{Liu(2020)}]{Liu2020a}
Liu, R. 2020, Research in Astronomy and Astrophysics, 20, 165

\bibitem[{Liu {et~al.}(2012)Liu, Berger, \& Low}]{Liu2012d}
Liu, W., Berger, T.~E., \& Low, B.~C. 2012, \apjl, 745, L21

\bibitem[{{Longcope} \& {Forbes}(2014)}]{Longcope2014a}
{Longcope}, D.~W., \& {Forbes}, T.~G. 2014, \solphys, 289, 2091

\bibitem[{{Low}(1996)}]{Low1996a}
{Low}, B.~C. 1996, \solphys, 167, 217

\bibitem[{Lugaz {et~al.}(2017)Lugaz, Farrugia, Winslow, Small, Manion, \&
  Savani}]{Lugaz2017}
Lugaz, N., Farrugia, C.~J., Winslow, R.~M., {et~al.} 2017, \apj, 848, 75

\bibitem[{{Moore} {et~al.}(2001){Moore}, {Sterling}, {Hudson}, \&
  {Lemen}}]{Moore2001a}
{Moore}, R.~L., {Sterling}, A.~C., {Hudson}, H.~S., \& {Lemen}, J.~R. 2001,
  \apj, 552, 833

\bibitem[{{Parenti}(2014)}]{Parenti2014a}
{Parenti}, S. 2014, Living Reviews in Solar Physics, 11, 1

\bibitem[{{Savcheva} {et~al.}(2012){Savcheva}, {van Ballegooijen}, \&
  {DeLuca}}]{Savcheva2012b}
{Savcheva}, A.~S., {van Ballegooijen}, A.~A., \& {DeLuca}, E.~E. 2012, \apj,
  744, 78

\bibitem[{{Sterling} \& {Moore}(2004)}]{Sterling2004}
{Sterling}, A.~C., \& {Moore}, R.~L. 2004, \apj, 602, 1024

\bibitem[{{Su} {et~al.}(2011){Su}, {Surges}, {van Ballegooijen}, {DeLuca}, \&
  {Golub}}]{Su2011a}
{Su}, Y., {Surges}, V., {van Ballegooijen}, A., {DeLuca}, E., \& {Golub}, L.
  2011, \apj, 734, 53

\bibitem[{Titov {et~al.}(2003)Titov, Galsgaard, \& Neukirch}]{Titov2003}
Titov, V.~S., Galsgaard, K., \& Neukirch, T. 2003, \apj, 582, 1172

\bibitem[{{Titov} {et~al.}(1993){Titov}, {Priest}, \& {Demoulin}}]{Titov1993a}
{Titov}, V.~S., {Priest}, E.~R., \& {Demoulin}, P. 1993, \aap, 276, 564

\bibitem[{{T{\"o}r{\"o}k} \& {Kliem}(2003)}]{Torok2003a}
{T{\"o}r{\"o}k}, T., \& {Kliem}, B. 2003, \aap, 406, 1043

\bibitem[{Tsap {et~al.}(2019)Tsap, Filippov, \& Kopylova}]{Tsap2019}
Tsap, Y.~T., Filippov, B.~P., \& Kopylova, Y.~G. 2019, \solphys, 294, 35

\bibitem[{{{\v S}vestka}(2001)}]{svestka2001a}
{{\v S}vestka}, Z. 2001, \ssr, 95, 135

\bibitem[{{Van Tend} \& {Kuperus}(1978)}]{vanTend1978a}
{Van Tend}, W., \& {Kuperus}, M. 1978, \solphys, 59, 115

\bibitem[{{Vr{\v s}nak} {et~al.}(2005){Vr{\v s}nak}, {Sudar}, \& {Ru{\v
  z}djak}}]{Vrvsnak2005a}
{Vr{\v s}nak}, B., {Sudar}, D., \& {Ru{\v z}djak}, D. 2005, \aap, 435, 1149

\bibitem[{{Xia} \& {Keppens}(2016)}]{Xia2016}
{Xia}, C., \& {Keppens}, R. 2016, \apj, 823, 22

\bibitem[{Yan {et~al.}(2020)Yan, Xue, Cheng, Zhang, Wang, Kong, Yang, Chen, \&
  Feng}]{Yan2020}
Yan, X., Xue, Z., Cheng, X., {et~al.} 2020, \apj, 889, 106

\bibitem[{{Zhang} {et~al.}(2001){Zhang}, {Dere}, {Howard}, {Kundu}, \&
  {White}}]{Zhang2001}
{Zhang}, J., {Dere}, K.~P., {Howard}, R.~A., {Kundu}, M.~R., \& {White}, S.~M.
  2001, \apj, 559, 452

\bibitem[{{Zhang} {et~al.}(2016){Zhang}, {Wang}, {Hu}, \& {Liu}}]{Zhang2016a}
{Zhang}, Q., {Wang}, Y., {Hu}, Y., \& {Liu}, R. 2016, \apj, 825, 109

\bibitem[{{Zhang} {et~al.}(2017){Zhang}, {Wang}, {Hu}, {Liu}, \&
  {Liu}}]{Zhang2017a}
{Zhang}, Q., {Wang}, Y., {Hu}, Y., {Liu}, R., \& {Liu}, J. 2017, \apj, 835, 211

\bibitem[{Zhang {et~al.}(2020)Zhang, Wang, Liu, Zhang, Hu, Wang, Zhuang, \&
  Li}]{Zhang2020}
Zhang, Q., Wang, Y., Liu, R., {et~al.} 2020, \apjl, 898, L12

\bibitem[{Zhang {et~al.}(2017)Zhang, Li, Zheng, Su, \& Ji}]{Zhang2017b}
Zhang, Q.~M., Li, T., Zheng, R.~S., Su, Y.~N., \& Ji, H.~S. 2017, \apj, 842, 27

\end{thebibliography}

\end{document}